\documentclass[%
reprint,
showpacs,
 amsmath,amssymb,
aps,
prb,
]{revtex4-1}

\usepackage{amsfonts}
\usepackage{amsmath, amssymb}
\usepackage{bm} % bold math
\usepackage{graphicx} % Include figure files
\usepackage{color}
\usepackage{ulem}
\graphicspath{{pics/}}

\newcommand{\veps}{\varepsilon}

\renewcommand{\Im}{\mbox{Im}}
\renewcommand{\Re}{\mbox{Re}}

\newcommand{\bee}{\begin{equation}}
\newcommand{\ene}{\end{equation}}

\newcommand{\bea}{\begin{eqnarray}}
\newcommand{\ena}{\end{eqnarray}}

\newcommand{\beg}{\begin{gather}}
\newcommand{\eng}{\end{gather}}

\newcommand{\dd}{\partial}

\newcommand{\ve}[1]{{\bf {#1}}}
\newcommand{\te}[1]{{\widehat{#1}}}

\begin{document}

%\preprint{APS/123-QED}

\title{Optical binding via surface plasmon polariton interference }

\author{Natalia Kostina$^1$}
\author{Aliaksandra Ivinskaya$^1$}
\author{Sergey Sukhov$^2$}
\author{Andrey Bogdanov$^1$}
\author{Ivan Toftul$^{1,3}$}
%\author{S.M.B. Blade-coating-guys}
\author{Manuel  Nieto-Vesperinas$^{4}$}
\author{Pavel Ginzburg$^{1,5}$}
\author{Mihail Petrov$^{1,3}$}
\author{Alexander Shalin$^1$}
\affiliation{$^1$ITMO University, Department of Nanophotonics and Metamaterials, Saint-Petersburg, 199034, Russia }

\affiliation{$^2$CREOL, University of Central Florida, Tampa, 32816-2700, USA}
\affiliation{$^3$Saint-Petersburg Academic University, Saint-Petersburg, 194021, Russia}
\affiliation{$^4$Instituto de Ciencia de Materiales de Madrid, Consejo Superior de Investigaciones Cientificas, Campus de Cantoblanco, Madrid E-28049, Spain}
\affiliation{$^5$ School of Electrical Engineering, Tel Aviv University, Tel Aviv, 6997801,  Israel}

%\affiliation{$^{1}$ITMO University, Birzhevaya liniya 14, 199034 St.-Petersburg, Russia}

\date{\today}% It is always \today, today,

\begin{abstract}

%Optical binding allows formation of mechanically stable nanoparticle configuration due to the interference of the field  scattered of  subwavelength nanoparticles. In this paper we consider optical binding  in the vicinity of metallic interface based on surface plasmon polariton interference. We show that the  interparticle distance at  equilibrium is governed by the surface plasmon wave vector rather than by the photon wavevector, allowing for optical binding at subwavelength distances. Moreover, the SPP-binding between two dipole nanoparticles occurs along the direction of their dipole moment, contrary to vacuum binding, where a stable configuration is observed for dipoles oriented perpendicularly to the field polarization direction.  

Optical binding allows creation of mechanically stable nanoparticle configurations owing to formation of self-consistent optical trapping potentials. While the classical diffraction limit prevents achieving deeply subwavelength arrangements, auxiliary nanostructures enable tailoring optical forces via additional interaction channels. Here, a dimer configuration next to metal surface was analyzed in details and the contribution of surface plasmon polariton waves was found to govern the interaction dynamics. It was shown that the interaction channel, mediated by resonant surface waves, enables achieving subwavelength stable dimers. Furthermore, the vectorial structure of surface modes allows  binding between two dipole nanoparticles  along the direction of their dipole moments, contrary to vacuum binding, where a stable configuration is formed in the direction oriented perpendicularly to the polarization of dipole moments. In addition, the  enhancement of optical binding stiffness for one order of magnitude was predicted owing to the surface plasmon polariton interaction channel. These phenomena pave a way for developing new flexible optical manipulators, allowing for the control over a nanoparticle trajectory on subwavelength scales and opens a room of opportunities for optical-induced anisotropic, i.e. with different periods along the field polarization as well as  perpendicular to it, organization of particles on a plasmonic substrate.
%AS:Optical binding allows creation of mechanically stable nanoparticle configurations owing to formation of self-consistent optical trapping potentials. While classical diffraction limit prevents achieving deeply subwavelength arrangements, auxiliary substartes enable tailoring optical forces via additional interaction channels. Here, dimer formation next to metal surfaces was analyzed in details and contribution of surface plasmon-polariton waves was found to govern the interaction dynamics. We show that the interparticle distance at equilibrium is governed by the surface plasmon waves rather than by the photons, allowing for optical binding at subwavelength distances. Moreover, the SPP-binding between two dipole nanoparticles occurs along the direction of their dipole moment, contrary to vacuum binding, where a stable configuration is observed for dipoles oriented perpendicularly to the field polarization direction. T

\end{abstract}

\pacs{}% PACS, the Physics and Astronomy
\maketitle
\section*{Introduction}
Light carries momentum which can influence  matter through optical forces enabling manipulation of micro and nanoscale objects\cite{Marago2013} and even atom ensembles\cite{Bloch2012}. The methods of optical tweezing\cite{Ashkin1970,Letokhov1968} rely on attraction of small objects to the regions of high field intensity. Spatially non-uniform intensity distributions used for positioning microobjects at predefined pattern can be achieved with a nanostructured environment or by interference of several beams. Yet, since  early years of optical tweezing experiments, it was discovered that several particles tend to self-organize under homogeneous illumination\cite{Burns1989, Burns1990}. This effect is referred to as transverse optical binding. The interference between incident and scattered light, owing to interaction with particles, results in the formation of a set of potential wells defining stable positions of particles. Optical binding  has been intensively studied both theoretically \cite{Depasse1994,Chaumet2001, Chaumet2010, Mazilu2012,Sukhov2015}  and experimentally \cite{Bowman2013, Wei2016, Chvatal2015, Demergis2012, Simpson2017} also as a prospective method for self-organization of particles. However, the strength of optical binding drops rapidly with  nanoparticle size  as the scattering efficiency decreases as $\sim R^6$, where $R$ is the nanoparticle radius. On the other hand, the viscous damping is also reduced for smaller particles, which makes the fluctuations and stochastic processes in liquids to be more influential. As a result, for reliable optical control of subwavelength nanoparticles, strong optical fields are required. In order to achieve this without strong heating of the surrounding media, it was suggested to use plasmonic structures, which can enhance optical fields locally.  The optical binding can be enhanced by  localized plasmon resonance in the nanoparticles\cite{Chaumet2001,Salary2016, VanVlack2011}. The localized plasmons can improve trapping efficiency at hot spots of corrugated metal\cite{Juan2011, Quidant2008}, or provide particle acceleration against beam direction in plasmonic V-grooves \cite{Shalin2013}. Three dimensional structures of plasmonic particles, such as metamaterials, can be also employed to trap or manipulate nanoparticles, e.g., for realization of optical pulling forces attracting nanoparticles to a light source \cite{Shalin2015,Bogdanov2015a}. In the context of optical binding flat metal surfaces also may be  very relevant. The excitation of propagating surface plasmon polartions (SPPs) and induced optical thermal forces are responsible for self-organization of micron size nanoparticles \cite{Garces-Chavez2006}. Moreover, the  direct momentum transfer from SPP to  micron size particles\cite{Volpe2006,Yuan2017} can be used for enhancing the optical forces near planar metallic surface, which can be used for sorting and ordering of nanoparticles \cite{Demergis2012, Yan2014,Min2013}. Recently, it was suggested that SPP modes can open a way for manipulating the optical forces acting on nanosize particles by the directional excitation of the propagating SPP modes\cite{Wang2014, Petrov2016,Ivinskaya2017}. \\
Here, we propose a new mechanism of transverse optical binding via excitation  of SPP modes ({\it SPP binding}) near a metallic planar interface. This mechanism is based on far-field interaction through the interference of SPP waves and is different to formation of resonant nanoparticle molecules due to their near field interaction \cite{Valdivia-Valero2012,Aunon2014,VanVlack2011,Salary2016}. Comparing  to common transverse binding in a free space ({\it photon binding}), the proposed approach has several advantages: i) it can enhance  the binding effect for  small nanoparticles due to resonant excitation of SPP modes; ii) the  distance between the bounded nanoparticles is defined by the SPP effective wavelength and, thus,  can be significantly smaller surpassing the diffraction limit; iii) the binding occurs in the direction of dipole polarization in accordance with the directivity of  SPP emission, which differs from the case of a free space binding, where stable configurations are formed in the  direction perpendicular to the dipole moments. In this paper, we  theoretically show how SPP-based transverse optical binding can bring new features to nanoparticle trapping and manipulation.

\section{Nanoparticles polarization near a substrate }      

\begin{figure}[h!]
\includegraphics[width=1\columnwidth]{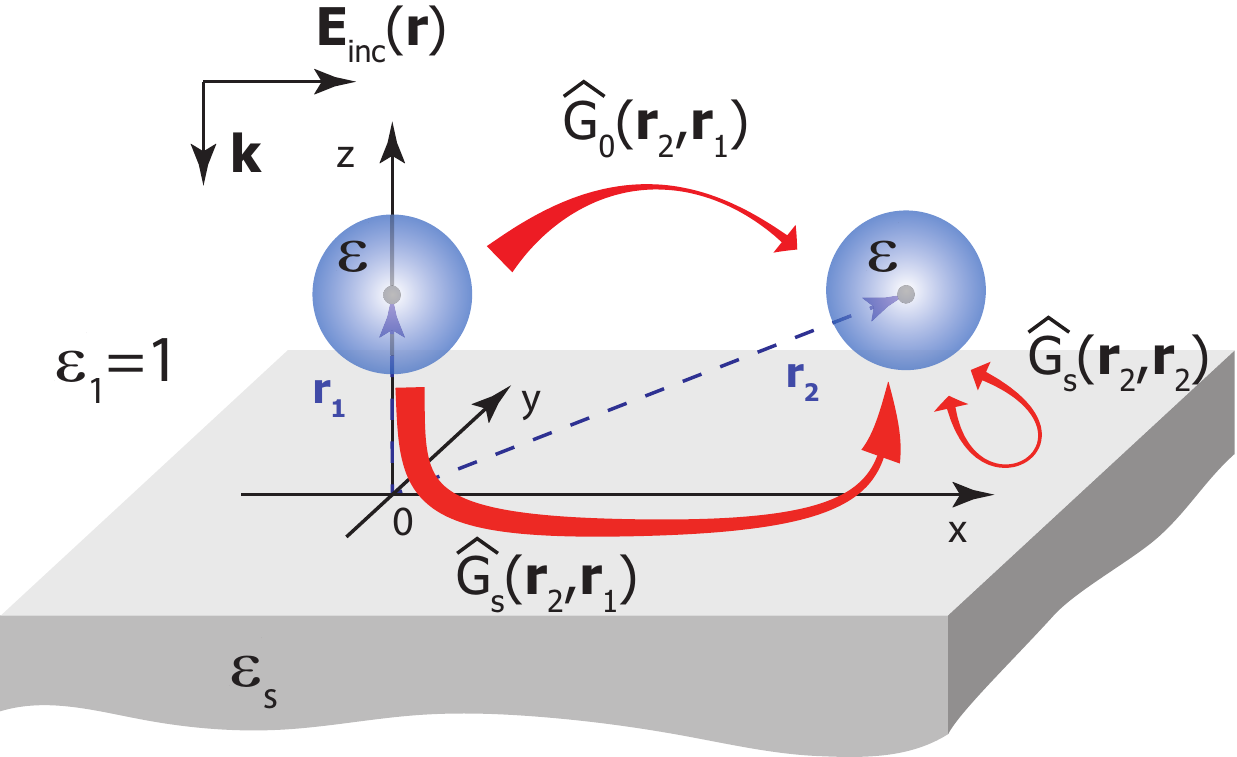}
\caption{  The scheme of the problem. Nanoparticles with  permittivity  $\veps$ are positioned at equal distances from a surface $z_1=z_2=z$. We  assume that the permittivity of the upper half-space equals to $\veps_1=1$.}
\label{Fig1}
\end{figure}

We consider two identical nanoparticles placed close to a planar metallic interface at coordinates $\ve r_1$, and $\ve r_2$ in the field of a normally incident plane wave (see  Fig.~\ref{Fig1}). We assume that nanoparticles have radius $R$ and  are made of dielectric material with permittivity $\veps$. In the dipole approximation  the radius of nanoparticles $R$ is much smaller than the typical scale of electric field variations. In this limit,  the  optical force acting on a  nanoparticle is given by the expression\cite{Chaumet2003} 
\begin{equation}
\label{DipoleForce}
{\bf F}=\dfrac12\Re\sum_i p_i^*{ \nabla} E_i({\bf r},\omega), 
\end{equation} 
where $E_i(\ve r,\omega)$ is the $i$-th component of a local field. 
%===============================FIGURE====================================================================
%===============================FIGURE====================================================================

The dipole moment of a nanoparticle $\ve p(\ve r)$ is defined  as ${\bf p(\ve r)}={\alpha}(\omega){\bf E}({\bf r})$, where  $\alpha(\omega)$ is the vacuum dipole polarizability corrected with account for retardation effects:
\bee
\dfrac{1}{\alpha}=\dfrac{1}{\alpha_0}-\dfrac{ik_0^3}{6\pi\veps_0 }, \ \alpha_0=4\pi\veps_0R^3\dfrac{\veps-\veps_1}{\veps+2\veps_1}, 
\ene
where $k_0$ is the wavevector in a free space, and $\veps_0$ is the vacuum permittivity.   
 The local electric field   includes  the incident plane wave, multiply rescattered field between particles via free-space and substrate  channels, and self-induced contribution of each particle through the reflection from the substrate. The local field is given by %in the position of the first and second dipole  can be expressed as follows:  

%\bea
%\label{System0}
%{\bf E}({\bf r}_1)={\bf E}_0({\bf r}_1)+\dfrac{k^2}{\veps_0}\te G_s(\ve r_1, \ve r_1)\ve p_1 +\dfrac{k^2}{\veps_0}\te G(\ve r_1, \ve r_2)
%\ve p_2,\\
%{\bf E}({\bf r}_2)={\bf E}_0({\bf r}_2)+\dfrac{k^2}{\veps_0}\te G_s(\ve r_2, \ve r_2)\ve p_2 +\dfrac{k^2}{\veps_0}\te G(\ve r_2, \ve r_1)
%\ve p_1.
%\ena
%
\bea
\label{System0}
{\bf E}({\bf r})={\bf E}_0({\bf r})+\dfrac{k_0^2}{\veps_0}\te G(\ve r, \ve r_1)\ve p_1 +\dfrac{k_0^2}{\veps_0}\te G(\ve r, \ve r_2)
\ve p_2.
%i=1,2\quad j=2,1 \nonumber
\ena
%   electric field generated by a dipole in this system can be expressed in terms of a Green's function $\ve E(\ve r)=k^2/\veps_0 \widehat{G}(\ve r, \ve r_0)\ve p$, where the Green's function can be decomposed into two components $\te G(\ve r, \ve r _0)=\te G_0(\ve r,\ve r_0)+\te G_s(\ve r, \ve r_0)$. The first component  $\widehat G_0$ is vacuum Green's function, and the second one $\widehat G_s$ is the scattered part \cite{Novotny}. 
%===============================FIGURE====================================================================
\begin{figure}[t!]
\includegraphics[width=0.7\columnwidth]{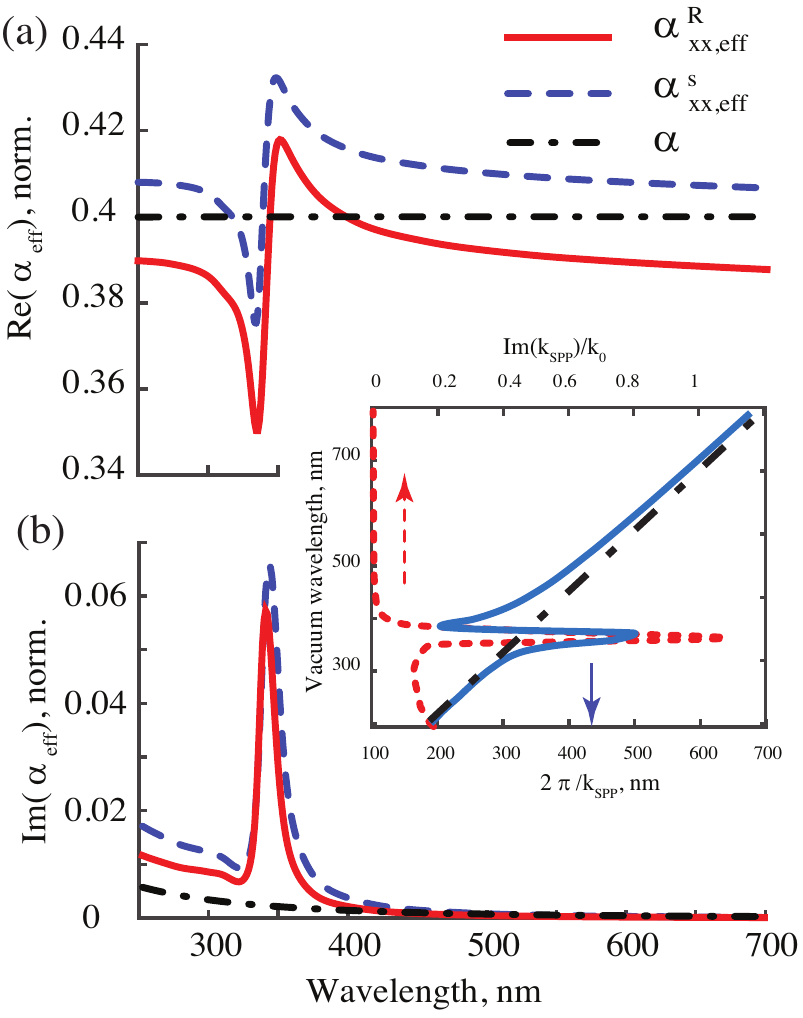}
\caption{ The spectrum of real (a)  and imaginary (b) parts of the $xx$-components of effective polarizability tensor $\hat\alpha^s_{eff}$ (blue dashed line) and $\hat\alpha^R_{eff}$ (red solid line) shown along with vacuum polarizability $\alpha$ (black dash-dot line). %The $xx$- components are shown along with the spectrum of vacuum polarizability $\alpha$. 
The polarizability is normalized over $4\pi \veps_0 R^3$, and calculated for a nanoparticle of radius $R=15$ nm with $\veps=3$ whose center is located above the surface at $z=20$ nm.  The inset shows the dependence of the SPP typical period denoted as $L_{SPP}=2\pi/k_{SPP}$ (upper $x$-axis), and imaginary part of SPP wavevector (lower $x$-axis) on the vacuum wavlength.        }
\label{Fig2}
\end{figure}
%===============================FIGURE====================================================================

Here, the first term in the right hand side is the amplitude of the external field, the second and third terms correspond to the field  generated by  the first and second  nanoparticles respectively.  The total  Green's function $\te G(\ve r, \ve r _0)=\te G_0(\ve r,\ve r_0)+\te G_s(\ve r, \ve r_0)$ is a sum of the scattered $\te G_s$ and vacuum $\te G_0$ components respectively \cite{Novotny2012}. One can simplify the consideration  if renormalizes the polarizability  tensor with respect to the self-action Green's function component $\te G_s(\ve r_i,\ve r_i)$ term. 
\bea
\label{System1}
\ve p_i=\te \alpha _{i,eff}^s \left({\bf E}_0({\bf r}_i) +\dfrac{k_0^2}{\veps_0}\te G(\ve r_i, \ve r_j)\ve p_j\right),\\
i=1,2\quad j=2,1. \nonumber
\ena
Here, we have introduced the effective polarizability tensor $\te \alpha_{i,eff}^s$ as follows: 
\bee 
\label{AlphaEff}
\widehat{\alpha}_{i, eff}^{s}({\ve r_i},\omega)=\alpha(\omega)\left(1-\alpha(\omega)\dfrac{k_0^2}{\veps_0}\widehat{G}_s({\ve r_i},{\ve r_i},\omega)\right)^{-1},\ \mbox{i=1,2}.
\ene
This tensor gives a correction of a vacuum polarizability $\alpha(\omega)$ with  accounting  for nanoparticle self-action through the substrate. This tensor is diagonal as $\te G_s(\ve r_i,\ve r_i)$ is diagonal in the case of a flat isotropic substrate \cite{Novotny2012}.

For the sake of simplicity, in the following we fix the position of the first particle in the origin of the coordinate system at $x_1=0$ and $y_1=0$, and will consider the force acting on the second particle only.  Computing the field at the point of the dipole according to the expression in Eq.~(\ref{System0}), one can achieve a system of equations for dipole moments $\ve p_1$ and $\ve p_2$ (see Appendix~\ref{AppI}), and,   in the special case of normal incidence of the plane wave, the expression for the dipole moments can be simplified even further: 
\begin{gather}
\ve p_i=\te \alpha _{i,eff}^{R} {\bf E}_0({\bf r}_i), \\
%\te \alpha _{i,eff}^{R} =\te \alpha _{1,eff}^{r}\left(1 +\dfrac{k^2}{\veps_0}\te G(\ve r_1, \ve r_2)\te \alpha _{2,eff}^s\right).
\widehat{\alpha}_{i, eff}^R({\ve r_i},\omega)=\alpha(\omega)\left(1-\alpha(\omega)\dfrac{k^2}{\veps_0}\widehat{G}_s({\ve r_i},{\ve r_i},\omega)-\right.\nonumber \\
\left.\dfrac{k^4}{\veps_0^2} \alpha (\omega) \te G(\ve r_i, \ve r_j)\te \alpha _{j,eff}^s(\ve r_j,\omega)\te G(\ve r_j, \ve r_i) \right)^{-1}\times \\
\left(1 +\dfrac{k^2}{\veps_0}\te G(\ve r_i, \ve r_j)\te \alpha _{j,eff}^s(\ve r_j,\omega)\right),\nonumber\\ 
{i=1,2\quad j=2,1}\nonumber.
\end{gather}

Now, the polarizability $\te \alpha^{R}_{eff}$ (see the Appendix~\ref{AppI} for the details) includes all the interaction channels: (i) the self-action of nanoparticle through the substrate, and (ii) the cross-action of two nanoparticle via vacuum and substrate. Moreover, it is worth mentioning that though the effective polarizability tensor $\te \alpha^s_{eff}$ is diagonal, the tensor $\te \alpha^{R}_{eff}$ is non-diagonal as the presence of the second nanoparticle does not preserve  translational symmetry of the system.

The excitation of SPP modes affects both the effective polarizability due to the substrate mediated self-action, and cross action of nanoparticles.  %The effective polarizability tensor is diagonal with two different components $\alpha_{xx, eff}^s=\alpha_{yy, eff}^s$ and $\alpha_{zz, eff}^s$. 
The spectra of real and imaginary parts of $xx$ - components of $\te \alpha_{eff}^{R}$ (solid line) and $\te \alpha_{eff}^{s}$ (dashed line) are plotted in Fig.~\ref{Fig2} for the case of silver substrate. The vacuum polarizability  $\alpha$ is also shown in the figure with dash-dot line. One can see that the effective polarisabilities have resonance at around 350 nm. From the inset of Fig.~\ref{Fig2}, one can see that this wavelength corresponds to SPP resonant excitation for silver/vacuum interface, which is defined by the condition  $\Re(\veps_s(\omega))+1=0$ and also corresponds to maximal  value of real part of SPP wavevector   $k_{SPP}=k_0\sqrt{\veps_s/(\veps_s+1)}$. In the inset the effective wavelength of SPP mode defined as $L_{SPP}=2\pi/k_{SPP}$ is also shown. The strong enhancement of the imaginary part of the effective polarizability is a sign of  strong rescattering of light  into the SPP mode. 

\section{Determining the stable  configurations }

\begin{figure*}[t]
\includegraphics[width=0.7\textwidth]{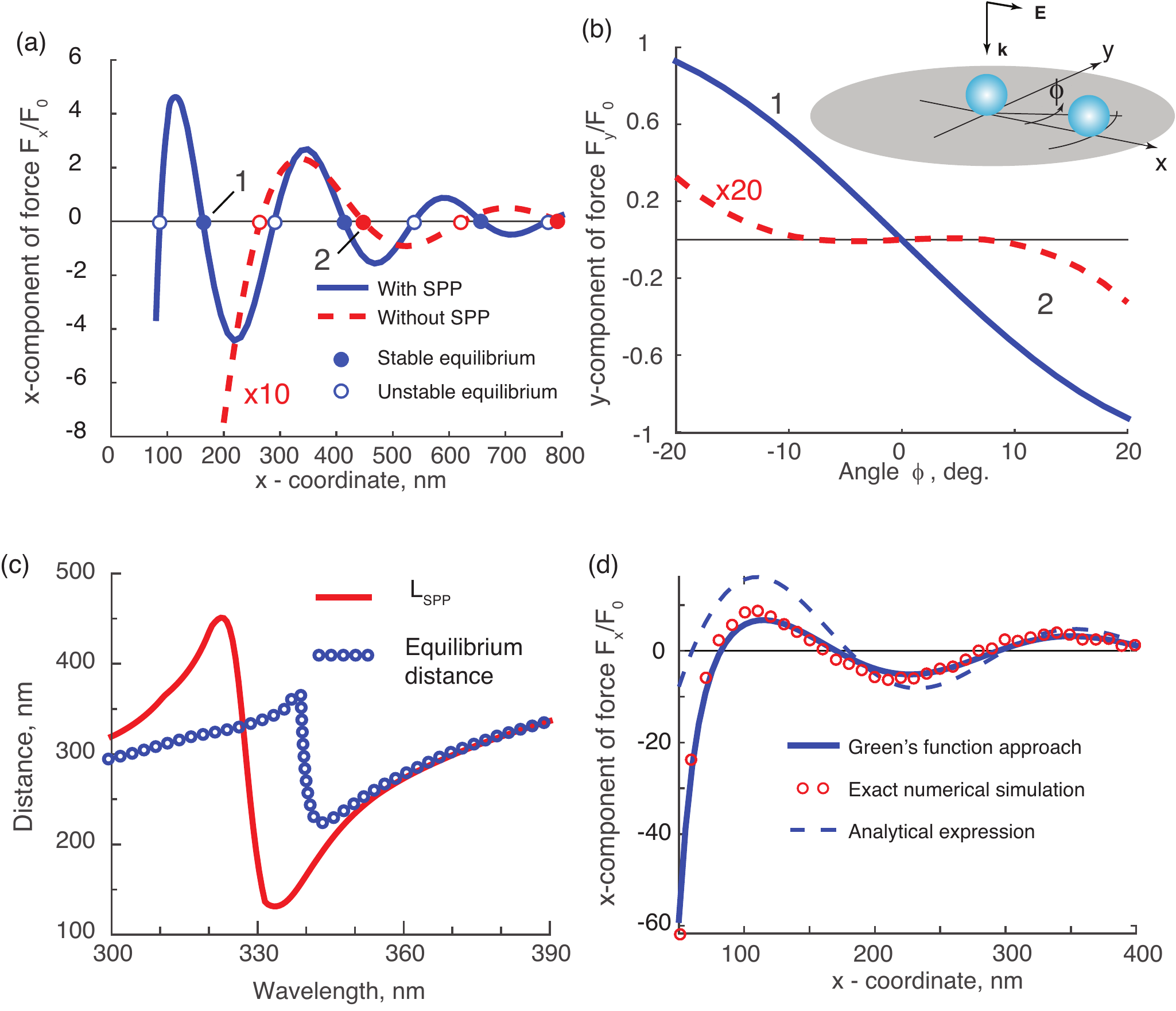}
\caption{(a) The $x$-component of optical force plotted along the $x$-axis coinciding with the direction of electric field polarization. The blue solid line denotes the force with account for all interaction channels. The red dashed line is for the interaction through the free space photons only. (b) The $y$-component of optical force in the direction perpendicular to $x$-axis, showing the stability of the binding position in the direction transverse to the $x$-axis. The blue solid line and red dashed line correspond to force calculated with or without account for SPP interaction channel for the equilibrium positions labeled 1 and 2 in a) correspondingly. The results are shown for the wavelength $\lambda=350$ nm.  (c) The distance between the stable equilibrium positions obtained from  (a) compared with the distance $L=2\pi/k_{SPP}$. (d) The comparison of the optical force calculated within Green's function approach (same as in Fig.\ref{Fig3} (a)) and calculated numerically with COMSOL Multyphysics package. The approximate analytical expression given by the 
Eq.~(\ref{ForceX_anal}) for SPP induced force is also shown with dashed line. All the results shown in the figure are computed for $R=15$ nm and $z=25$ nm. }
\label{Fig3}
\end{figure*}

By determining the dipole moments of nanoparticles, one can calculate the optical force acting on each nanoparticle using the expression \eqref{DipoleForce} (see the details in the Appendix Eq.~\eqref{Fx}). In the following, we will refer to the optical force acting on the second nanoparticle only, fixing the first nanoparticle in the coordinate origin. In order to find the equilibrium positions of nanoparticle, we plot the dependence of the $x$-component of the optical force as a function of interparticle distance along $x$-axis as shown in Fig.~\ref{Fig3} a). The force is normalized over the optical pressure force  acting on the same nanoparticle in vacuum $F_0=1/2k|E_0|^2\Im(\alpha(\omega))$, where $|E_0|$ is the amplitude of the incident plane wave. One can see that the force changes at particular point on the  $x$-axis denoting the equilibrium positions points. These points can be stable along $x$ if the force is restoring (shown with solid  circles, i.e. point 1), and unstable otherwise (shown with white filled circles). One should also note that when nanoparticles are close to each other the force goes to minus infinity, until  the nanoparticles touch each other. However this case is out of the scope of the present paper.  
%===============================FIGURE====================================================================

%===============================FIGURE====================================================================

To identify the role of  plasmons in the interaction force we have excluded SPP contribution from the Green's function by integrating over the free space modes only in the spectral representation (see Appendix \ref{GreensSubs}).  One can see that in the absence of SPPs the interaction force becomes one order of magnitude weaker, and the period between stable positions is significantly enlarged being defined by the vacuum wavelength and photons interference. Moreover, the  equilibrium points shown with blue circles are stable both  along $x$ and $y$ directions making them globally stable, which does not happen in case of photon binding. To illustrate this, we plotted the $F_y$ force (see Fig.~\ref{Fig3} b)) as function of the transverse angle $\phi$ (see the inset in Fig.~\ref{Fig3}) in the vicinity of points of stable  equilibrium positions. In Fig.~\ref{Fig3} (c) one can see the dependence of the binding length on the vacuum wavelength shown along with the period of SPP wave, which is equal to $L_{SPP}$. One can see that the binding distance is fully defined by the period of the SPP wave when the excitation condition is fulfilled, thus, providing the binding at distances   significantly shorter than vacuum wavelength.  This also strongly differs from work of Salary et. al. \cite{Salary2016}, where the optical forces between two nanoparticles over metallic substrate were considered  in the regime, when the interaction force is mainly defined by the near-field components. 
 %===============================FIGURE====================================================================
\begin{figure}[h!]
\includegraphics[width=1\columnwidth]{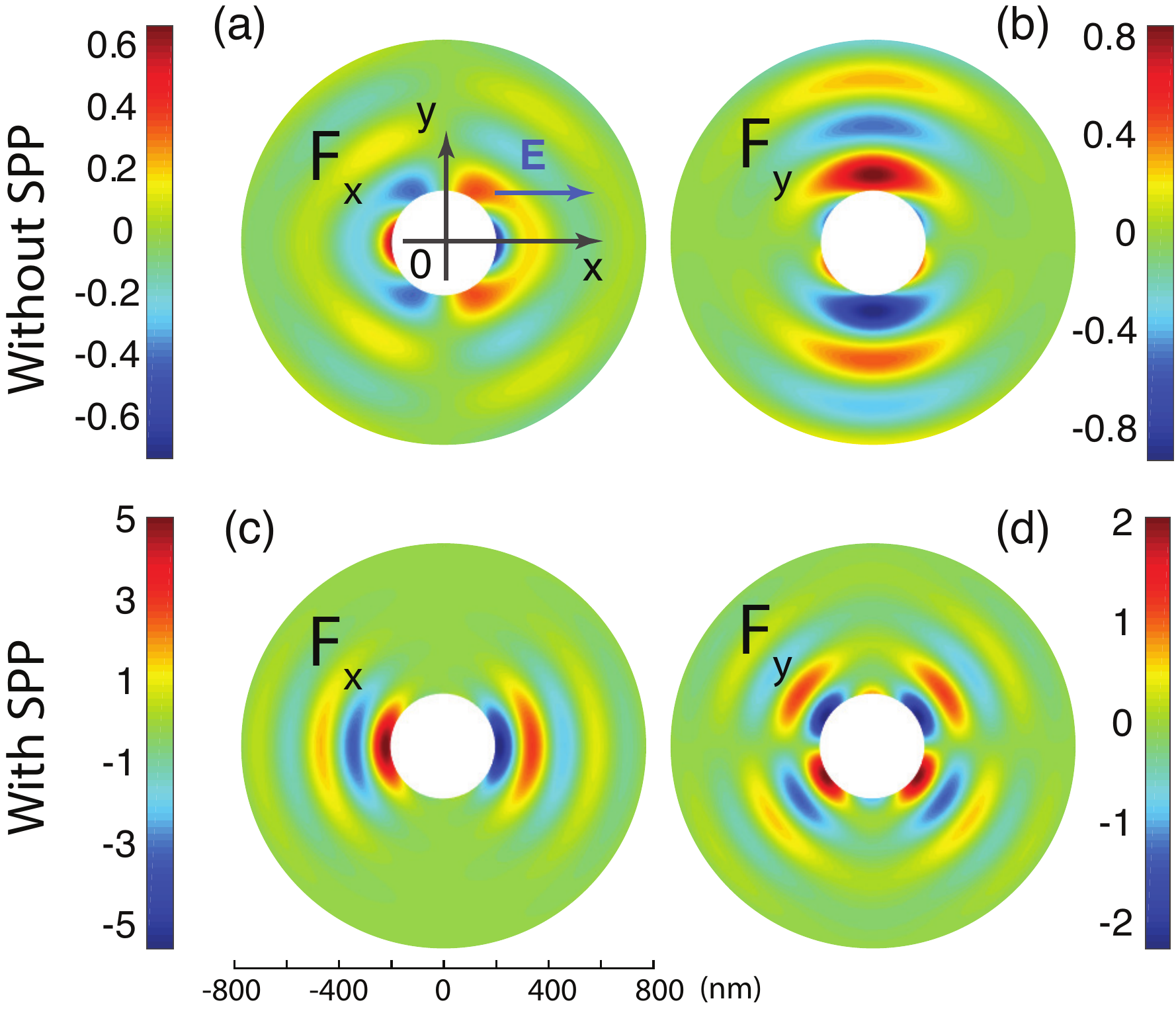}
\caption{ The two-dimensional maps showing the $x$ and $y$ forces for SPP and photon binding. The different directionality of the scattering pattern is responsible for  different geometry of stable equilibrium positions. The results are shown for wavelength $\lambda=350$ nm and $z=25$ nm. }
\label{Fig4}
\end{figure}
%===============================FIGURE====================================================================

In order to support the  results discussed above, we have performed numerical simulations in COMSOL Multyphysics package (see Fig.~\ref{Fig3} (d), scatter line)  and showed good correspondence with the obtained solution based on Green's function approach (see Fig.~\ref{Fig3} (d) solid line). The Green's function approach shows good agreement with numerical results also at the distances  comparable to nanoparticle size. Moreover, basing on Green's function formalism we have derived the approximate expression for the contribution of SPP mode into the optical force  (see  Appendix~\ref{ForceAnal} for the details):

\bea
%F_x\approx \pi |p_x|^2 \Re\left[\dfrac{(k^*)^3 (k_{1z}^*)^2k_{2z}^*}{k_0^2(\veps_1-\veps_2)}\left(H_1^{(1)}(k^* x)-\dfrac{J_2(k^* x)}{(k^*x)^2}\right)\exp(ik_{1z}^*z) \right].\nonumber
F_x\approx \pi |p_x|^2 \Re\left[\dfrac{k_{SPP}^3 k_{1z}^2k_{2z}}{k_0^2(1-\veps_s)}\right.\times \nonumber \\
\left.H_1^{(1)}(k_{SPP} x)\exp(-\Im(k_{1z})z) \right].
\label{ForceX_anal}
\ena

Here $k_{1z}=\sqrt{k^2_0-k_{SPP}^2}$, and $k_{2z}=\sqrt{\veps_s k^2_0-k_{SPP}^2}$ are $z$-components of SPP wavevector in  the upper half-space and in the substrate correspondingly, $H_1^{(1)}(q)$ is the first order Hankel function of the first kind. The derived expression very illustratively shows the origin of the SPP mode: the  Hankel function describes the SPP mode excited by a dipole and propagating over a flat surface. Its zeros define the  equilibrium positions of the nanoparticle. The $z$-components of the wavevector is complex since SPP is a localized wave, thus, the exponent in  Eq.~(\ref{ForceX_anal}) shows that strength of dipole-SPP coupling decays.  
    
%These points alternate with period defined by the SPP wavevector. 
%The positive derivative in case of SPP binding provides the restoring properties of optical force when nanoparticle is shifted perpendicularly to $y$-axis. 
One needs to stress, that the transverse binding in vacuum does not  not provide stable equilibrium positions along $x$-axis \cite{Dholakia2010}. This difference of SPP and photon binding can be understood through the  difference in the scattering diagrams of SPPs and photons. This is illustrated  in Fig.\ref{Fig4} where two-dimensional maps of $x$  and $y$ force components are plotted. The photon binding is well-known to have stable configuration perpendicular to the field polarization direction in accordance with the dipole emission pattern (see Fig.~\ref{Fig4} (a, b)). The SPP binding, on the contrary, has stable configurations along the polarization direction, in which preferable excitation of SPP modes occurs (see Fig.~\ref{Fig4} (c, d)). It is also worth noting that the amplitudes of lateral forces are several times  higher when SPP modes affect binding. 

\begin{figure}[h!]
\includegraphics[width=1\columnwidth]{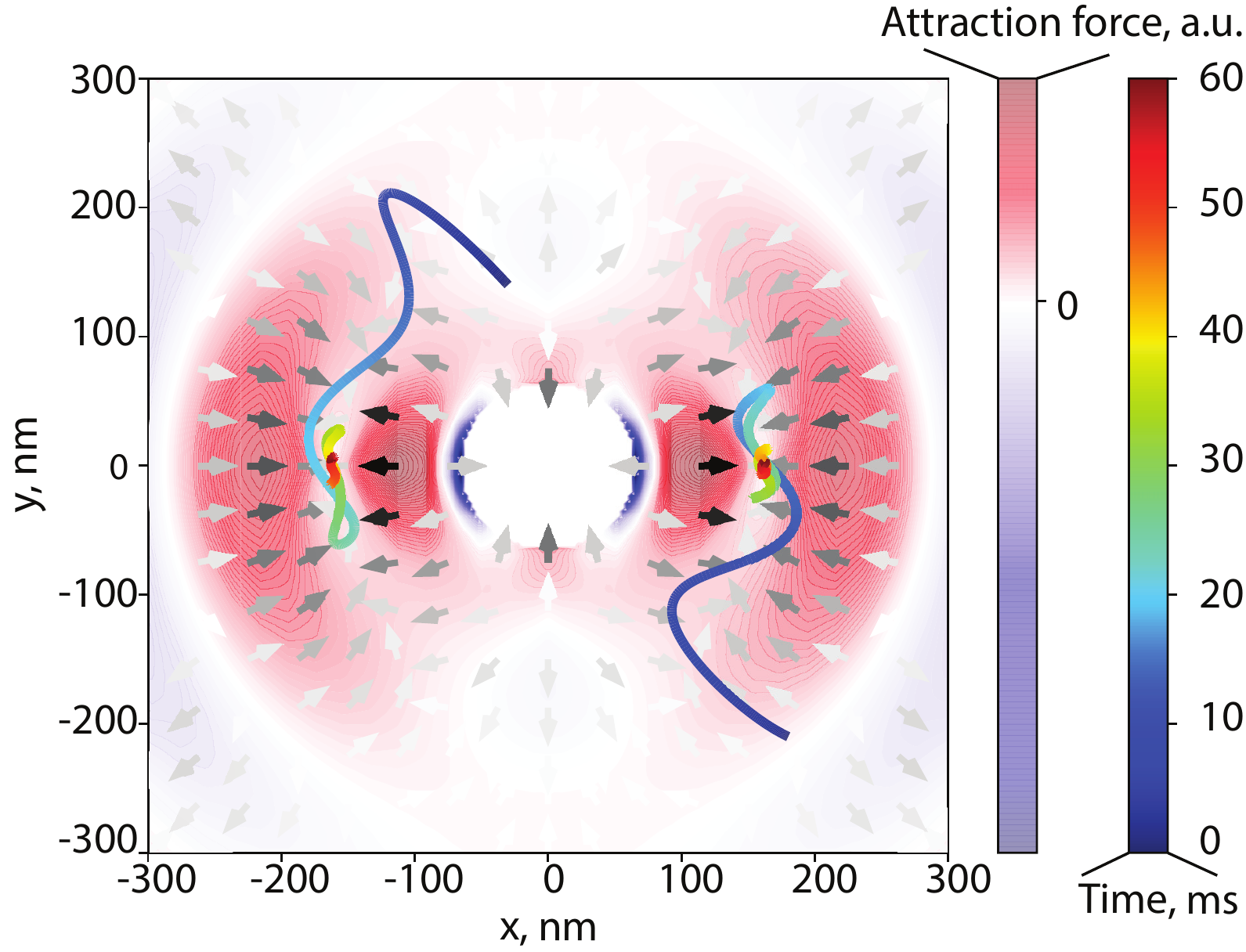}
\caption{The dynamics of the second nanoparticle motion. The first nanoparticle is fixed at the origin. The colour of the trajectory line denotes the time elapsed since the beginning of motion. The arrows show the force field: the darker are the arrows the stronger is the optical force. The colour map at the background shows the force which attracts or repulses nanoparticle to/from the equilibrium positions at $y=0$ and $x=\pm 175$ nm. The intensity of the red colour gives the strength of nanoparticle attraction, while blue shows repulsion of the nanoparticle. The parameters of computation are $R=15$ nm, $z=25$ nm, $\veps=3$. The laser intensity was taken $5\cdot 10^5$ W/m$^2$, and the dimensionless damping factor $\gamma=0.015$ (see Appendix~\ref{App:dynamics} for the details of the simulation method). }
\label{Dynamics}
\end{figure}

We illustrate the character of SPP binding by calculating the dynamics of the second nanoparticle motion in the force field of the first nanoparticle, which is fixed at the origin of the coordinates. We consider only two-dimensional  motion of the nanoparticle, keeping the $z$-coordinate  to be constant. The dynamics is obtained through direct solution of  equations of motion under the external optical force with the account for viscous damping. The details are discussed in Appendix~\ref{App:dynamics}.   
Two typical trajectories are shown in Fig.~\ref{Dynamics} for two different sets of initial coordinates of the second nanoparticle. The color map shows the intensity of nanoparticle attraction to the equilibrium positions along $x$-axis. The arrows show the force field, while the lines show the trajectories with colour changing from blue to red while time elapses.   One can see that the nanoparticle actively tends to set the position along the $x$-axis where the binding force is the strongest.   

%===============================FIGURE====================================================================
%===============================FIGURE====================================================================

\begin{figure}[h!]
\includegraphics[width=1\columnwidth]{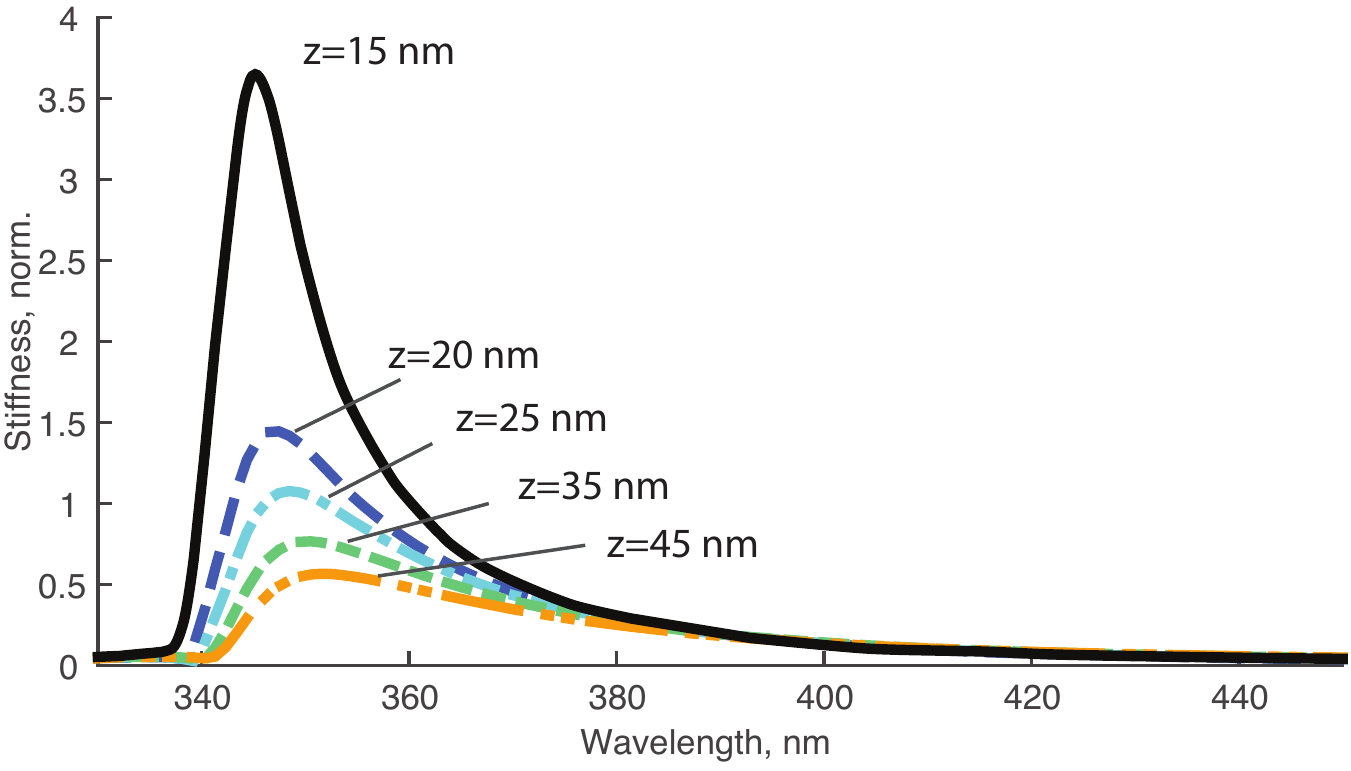}
\caption{ The  stiffness $\kappa_x$ in units of $\kappa_0=F_0/R$ as a function of the excitation wavelength. The spectra are shown for different distances $z$ from the nanoparticle center to the  surface. Nanoparticle radius $R=15$ nm.}%Inset: the binding length versus the  vacuum wavelength, and SPP period.  }
\label{Fig5}
\end{figure}

  The important parameter, which  characterizes the stability of the equilibrium states, is the stiffness of the trap. At the equilibrium positions the total optical force is zero, but when shifted from the stable positions the nanoparticles undergo action of a restoring force, which is locally proportional to the amplitude of the displacement $F_{r}=-\kappa_x \Delta x$, with the parameter $\kappa_x$ characterizing the stiffness of the system along the $x$-direction. However, this approximation of the restoring force only applies to the gradient component of the optical force. Indeed,  we consider the nanoparticles significantly smaller than the wavelength, that results in low and non-resonant at the wavelengths imaginary part of the polarizability $\Im (\alpha_{eff})\ll\Re(\alpha_{eff})$, as $ \Im (\alpha_{eff}) \simeq (R^6/\lambda^3)$, and 
$ R<<\lambda$ (see Fig.~\ref{Fig2}). Thus, the radiation force, which is proportional to imaginary part of the polarizability, can be neglected (see Appendix~\ref{App:conserve}).    The stiffness in the considered system strongly depends on the mechanism of the  nanoparticles interaction, and, as can be seen from Fig.~\ref{Fig3}, it is much higher when the plasmon interaction is enabled. We have plotted (see Fig.~\ref{Fig5}) the spectral dependence of stiffness parameter $\kappa_x$ calculated at the first equilibrium point, labeled  by point 1 in Fig.~\ref{Fig3} (a). To avoid the  dependence of stiffness on the illumination intensity, we have normalized it to the magnitude $\kappa_0=F_0/R$, which is the stiffness of a system where the vacuum pressure force $F_0$  can be restored when nanoparticle is  displaced for one radius from its equilibrium position. One can see that the stiffness has a strong resonant behaviour, which corresponds to the excitation of SPP modes at wavelengths longer than 350 nm. With the increase of the distance from nanoparticle center to the surface the stiffness rapidly drops, as the coupling with the SPP mode decreases.

From Fig.~\ref{Fig5} one can see that the spectral maximum of  stiffness depends on the height from the surface.   This spectral dependence can be understood better by analyzing the analytical expression Eq.~(\ref{ForceX_anal}). In the case of neglible losses in the substrate, one can get a simple expression for the stiffness at the first stable equilibrium point (see Appendix~\ref{ForceAnal} for more details): 

\begin{gather}
\label{AppStiff}
\kappa_x\approx \pi |p_x|^2 \dfrac{( k_{SPP})^3 | k_{1z}|^2| k_{2z}|}{k_0^2(1-\veps_s)}Y_2(q_{1})\exp(-| k_{1z}|z).
\end{gather}
Here $Y_2(q)$ is the cylindrical Webber function of the second order, and  $q_1$ is the first positive root of $Y_1(q_1)=0$. Note that in the regime of SPP excitation without ohmic losses the SPP wavevector can be in a range from $k_0$ to $+\infty$ when $1+\veps_s\rightarrow 0$. With that the expression Eq.~(\ref{AppStiff}) goes to zero in both limiting cases:
  $$\kappa_x\xrightarrow[ k_{SPP}\rightarrow 0, \infty]{}0,$$  
    which implies that the stiffness reaches its maximum at some particular wavelength. This wavelength can be defined for each  given distance over substrate $z$. The maximal stiffness  can be achieved close to SPP resonance when SPP wavevector equals to $\tilde k\approx6/z\gg k_0$. Then,  the maximal stiffness at the $n$-th equilibrium position decreases with distance to substrate as $z^{-6}$ (see Eq.~(\ref{AppMaxStiff})): 
\bee
\kappa_{x,n}\sim |p_x|^2 \left(\dfrac1z\right)^6\dfrac{1}{k_0^2}Y_2(q_{2n+1}).
\ene
 
%===============================FIGURE====================================================================
%===============================FIGURE====================================================================

%-----------Section----------------------------------------------------------------------------------------

\section{Conclusion}

In this work we consider  transverse optical binding based on surface plasmon polariton interference. We show that two nanoparticles placed in the vicinity of a plasmonic interface can form a stable bound dimer with binding length defined by the SPP wavelength. This allows formation of the dimers with interparticle   distance significantly shorter than the free-space wavelength suppressing the diffractional limit.  The binding states are formed along the direction of the incident field polarization, which on the contrary tothe photon binding, where the stable bound states are formed perpendicular to the polarization direction. The excitation of SPP modes also enhances the amplitude of the binding forces, resulting in  resonant enhancement of the trap stiffness.  

\newpage 

\appendix 
%-----------Section----------------------------------------------------------------------------------------
\section{Calculation of a binding force}
\label{AppI}

%\bee 
%{\bf F}=\dfrac12\Re\sum_i p_i^*\nabla E_i({\bf r},\omega).
%\ene
The force $\ve F_2(\ve r_2)$ acting on the second nanoparticle is calculated as 
\bee
\label{eq:A_Force}
\ve F_2(\ve r_2)=\dfrac12\Re \left[ \sum p_{2i}^*(\ve r_2)\nabla  E_{i}(\ve r_2)\right]. 
\ene
The introduction of the effective polarizability parameters significantly simplifies the formula  for the electric field in the center of the second nanoparticle $\ve E_2$:
\bee
\label{eq:A_Field}
{\bf E}({\bf r}_2)={\bf E}_0({\bf r}_2) +\dfrac{k_0^2}{\veps_0}\te G_s(\ve r_2, \ve r_2) \ve p_2+ \dfrac{k_0^2}{\veps_0}\te G(\ve r_2, \ve r_1) \ve p_1.
\ene
The dipole moments then can be expressed as: 
\begin{gather}
\label{System1}
\ve p_1=\te \alpha_{1,eff}^s\ve E (\ve r)= \nonumber \\ 
\te \alpha _{1,eff}^s \left({\bf E}_0({\bf r}_1) +\dfrac{k_0^2}{\veps_0}\te G(\ve r_1, \ve r_2)\ve p_2\right),\\
\ve p_2=\te \alpha _{2,eff}^s \left({\bf E}_0({\bf r}_2) +\dfrac{k_0^2}{\veps_0}\te G(\ve r_2, \ve r_1)\ve p_1\right),\\
\label{AlphaEff}
\widehat{\alpha}_{i, eff}^s({\ve r_i},\omega)=\alpha(\omega)\left(1-\alpha(\omega)\dfrac{k_0^2}{\veps_0}\widehat{G}_s({\ve r_i},{\ve r_i},\omega)\right)^{-1},\\
i=1,2.\nonumber
\end{gather}%\bee 
%\ene
Let us solve this equation system. After some manipulations one can get 
\begin{widetext}
\bea
\label{p1_eff}
\ve p_1=\te \alpha _{1,eff}^s \left({\bf E}_0({\bf r}_1) +\dfrac{k_0^2}{\veps_0}\te G(\ve r_1, \ve r_2)\left(\te \alpha _{2,eff}^s \left[{\bf E}_0({\bf r}_2) +\dfrac{k_0^2}{\veps_0}\te G(\ve r_2, \ve r_1)\ve p_1\right]\right)\right),\\
\ve p_1=\te \alpha _{1,eff}^s \left({\bf E}_0({\bf r}_1) +\dfrac{k_0^2}{\veps_0}\te G(\ve r_1, \ve r_2)\te \alpha _{2,eff}^s {\bf E}_0({\bf r}_2)\right) +\dfrac{k_0^4}{\veps_0^2} \te \alpha _{1,eff}^s \te G(\ve r_1, \ve r_2)\te \alpha _{2,eff}^s\te G(\ve r_2, \ve r_1)\ve p_1,
\ena
\end{widetext}

The last expression in (\ref{p1_eff}) can be simplified even further, if one renormalizes the effective polarizability tensor with account for nanoparticle cross action:

\begin{widetext}
\bee 
\label{AlphaEff2}
\widehat{\alpha}_{i, eff}^r({\ve r_i},\omega)=\alpha_i(\omega)\left(1-\alpha_i(\omega)\dfrac{k^2}{\veps_0}\widehat{G}_s({\ve r_i},{\ve r_i},\omega)-\dfrac{k^4}{\veps_0^2} \alpha _{i} \te G(\ve r_i, \ve r_j)\te \alpha _{j,eff}^s\te G(\ve r_j, \ve r_i) \right)^{-1},\ {i=1,2\quad j=2,1} .
\ene
\end{widetext}
Here the self-action Green's function $\te G_s(\ve r_i, \ve r_i)$ contains the scattered part only, whereas the cross-action part $\te G(\ve r_i,\ve r_j)=\te G_0(\ve r_i,\ve r_j)+\te G_s(\ve r_i,\ve r_j)$ include both vacuum and scattered parts determining the cross-interaction through vacuum and via substrate  respectively. The final expression for the dipole moment will be as follows:

\begin{gather}
\ve p_i=\te \alpha _{i,eff}^r \left({\bf E}_0({\bf r}_i) +\dfrac{k^2}{\veps_0}\te G(\ve r_i, \ve r_j)\te \alpha _{j,eff}^s {\bf E}_0({\bf r}_j)\right),\\
%\ve p_2=\te \alpha _{2,eff}^r \left({\bf E}_0({\bf r}_2) +\dfrac{k^2}{\veps_0}\te G(\ve r_2, \ve r_1)\te \alpha _{1,eff}^s {\bf E}_0({\bf r}_1)\right)
i=1,2 \ \ j=2, 1 
\end{gather}

The case of normal plane wave  incidence on a planar substrate, when the nanoparticles are located at the same height above the surface (see Fig.\ref{Fig1}) is of a particular interest. In this case the external electric field $\ve E_0$ is equal in the centres of both nanoparticles, and, thus, the dipole moment  has very simple form:
   
\begin{gather}
\label{alphaR}
\ve p_{i}=\te \alpha _{i,eff}^{R} {\bf E}_0({\bf r}_{i}), \\
\te \alpha _{i,eff}^{R} =\te \alpha _{i,eff}^{r}\left(1 +\dfrac{k^2}{\veps_0}\te G(\ve r_{i}, \ve r_{j})\te \alpha _{j,eff}^s\right),\\
i=1,2\quad j=2,1. \nonumber
\end{gather}

The optical force component, then, can be calculated as
\begin{gather}
 F_{2x}(\ve r_2)=\dfrac12\Re \left[ \sum_{n=x,y,z} p_{2n}^*(\ve r_2)\partial_{x_2}  E_{n}(\ve r_2)\right]= \nonumber \\
  \dfrac12\Re \left[ \sum_{n=x,y,z} p_{2n}^*(\ve r_2)\left(\partial_{x_2}  E_{0n}(\ve r_2)+ \dfrac{}{}\right.\right. \nonumber \\ 
\left.\left.  \dfrac{k_0^2}{\veps_0} \sum_{m=x,y,z} \partial_{x_2'}G_{s,nm}(\ve r_2',\ve r_2)p_{2m}\right.\right.  \nonumber \\
\left.\left.+\dfrac{k_0^2}{\veps_0} \sum_{m=x,y,z} \partial_{x_2}G_{nm}(\ve r_2,\ve r_1)p_{1m}\right)\right]. \label{Fx}
\end{gather}

The $y$ and $z$ components can be calculated  with the same expression (\ref{Fx}) by substituting the partial derivative with $\partial_y $ and $\partial_z$ correspondingly. 

%-----------Section----------------------------------------------------------------------------------------
\section{Green's function }
\label{GreensSubs}
The Green's function tensor of a two half-spaces with permittivities $\veps_1$ (for $z>0$) and  $\veps_2$ (for $z\le0$) can be expressed in the cylindrical coordinates through the reciprocal representation in wave-vector space\cite{Novotny2012} (for $z>0$): 

\begin{gather}
\te{G}(\rho,\varphi,z>0)=\dfrac{i k_1}{8 \pi}\int_0^{\infty}\widehat M(s, \rho, \varphi) \exp\left(is_{z1} z\right) ds,
\end{gather}

where $k_1$ is the wavevector in the upper space, and  $s=k_r/k_0$ and $s_{z1}=k_{z1}/k_0$ are the radial and $z$-components of the dimensionless wavevector  normalized over the wavevector in the free space. 

\begin{gather}
\te M(s, \rho, \varphi)= \left(
\begin{tabular}{ccc}
$m_{xx} $& $m_{xy}$ & $m_{xz}$ \\
$m_{yx}$ & $m_{yy}$ & $m_{yz}$ \\
$m_{zx}$ & $m_{zy}$ & $m_{zz}$ 
\end{tabular}
\right)\\ 
m_{xx}=\dfrac{s}{s_{z1}}r_s f(s,\rho, \varphi)-s s_{z1}r_p g(s,\rho,\varphi), \nonumber \\
m_{yy}=\dfrac{s}{s_{z1}}r_s g(s,\rho,\varphi)-s s_{z1}r_p f(s,\rho,\varphi), \nonumber \\
m_{zz}=2\pi J_0(s\rho)r_p \dfrac{s^3}{s_{z1}}, \nonumber \\
m_{xy}=m_{yx}=\dfrac{(r_s+s_{z1}^2r_p)}{ss_{z1}} h(s,\rho,\varphi), \nonumber \\
m_{xz}=-m_{zx}=-s r_p t(s,\rho,\varphi), \nonumber\\
m_{yz}=-m_{zy}=-s r_p w(s,\rho,\varphi),\nonumber
\end{gather}
where the functions $f(s,\rho,\varphi),g(s,\rho,\varphi), h(s,\rho,\varphi)$, $t(s,\rho,\varphi), w(s,\rho,\varphi)$ can be expressed:
\begin{gather}
f(s,\rho,\varphi)=2\pi\left(\sin^2(\varphi) J_0(s\rho)+\dfrac{J_1(s\rho)}{s\rho}\cos(2\varphi)\right),\nonumber \\
g(s,\rho,\varphi)=2\pi\left(\cos^2(\varphi) J_0(s\rho)-\dfrac{J_1(s\rho)}{s\rho}\cos(2\varphi)\right), \nonumber \\
h(s,\rho,\varphi)= \pi s^2J_2(s\rho)\sin(2\varphi),\\
t(s,\rho,\varphi)= 2\pi i\ s J_1(s\rho)\cos(\varphi),\nonumber \\
w(s,\rho,\varphi)= 2\pi i\ s J_1(s\rho)\sin(\varphi). \nonumber
\end{gather}
  
 Here $J_n(z)$ is the first kind Bessel function of the order $n$.
%------------------Anaytical Force--------------------------------------- 
\section{Analytical expression}
\label{ForceAnal}

Here we analyze the $x$-component of the optical force acting on nanoparticles when normal incident light is poralized along the $x$-axis. This is the case considered in Fig.~\ref{Fig3}. According to Eq.(\ref{DipoleForce}) the expression for the force will be as follows: 

\bee
 F_x=\dfrac12\Re\left(\sum_i p_i(\ve r)^*\partial_x E_i(\ve r_i)\right).
 \label{Force_X}
\ene 
In order to get a simple analytical result showing all the key features of the SPP-assisted force, we will take into account that  the effective $\hat \alpha^R$ tensor has diagonal domination, which implies that $\alpha_{ii}^R\gg\alpha_{ij}^R,\ \ i\neq j$.

The the expression (\ref{Force_X}) can be simplified:
\bee
F_x=\dfrac12\Re ( p_x^*\partial_x E_{x}^s).
\ene
The electrical field generated by the dipole at the distance $\ve r$ from the first nanoparticle can be expressed through the Green's function $ E_{x}^s(\ve r)=4\pi k_0^2 G_{s,xx}(\ve r,0)p_x$. 

Then we have an expression for the lateral component of the optical force written in a very simple form: 
\bee
\label{Force_X2}
F_x=2\pi k_0^2 |p_x|^2\Re \left(\partial_x G_{s,xx}(\ve r,0)\right)
\ene

The Greens function is expressed through the integral 

\begin{gather}
G_{s,xx}(x,y,z)=G_{s, xx}(\rho, \phi, z)=\frac{ik_1}{8\pi^2}\times\\ 
\int_0^{\infty}m_{xx}(\rho,s)\exp(is_{1z}z)ds, \nonumber\\
m_{xx}=s r_s(s)\dfrac{a_1(\rho,\phi)}{s_{1z}}-s s_{1z} r_p(s)a_2(\rho,\phi),\nonumber \\
a_1=2\pi\left(\sin(\phi_0)^2 J_0(s \rho)+\dfrac{J_1(s \rho)}{s \rho}\cos(2\phi_0)\right),\nonumber \\
a_2=2\pi\left(\cos(\phi_0)^2J_0(s \rho)-\dfrac{J_1(s\rho)}{s\rho}\cos(2\phi_0)\right),\nonumber \\
s_{1z}=\sqrt{1-s^2} \quad s_{2z}=\sqrt{\veps_2-s^2}.\nonumber 
\end{gather}

Here we use the same notation as in Appendix~\ref{GreensSubs}. We are interested only in the component containing $r_p$ term as only it gives rise to SPP response, and also we  put $\phi=0`$. Then,

\begin{gather}
G_{s,xx}(\rho, \phi, z)=\frac{ik_1}{8\pi^2}\int_0^{\infty}m'_{xx}(\rho,s)\exp(is_{1z}z)ds,\\
m'_{xx}=-s s_{1z} r_p(s)a_2(s,\rho),\nonumber \\
a_2=2\pi\left(J_0(s \rho)-\dfrac{J_1(s \rho)}{s\rho}\right).\nonumber 
\end{gather}

Next, we have 
 
\begin{gather}
G_{s, xx}(\rho, \phi, z)=-\frac{ik_1}{4\pi}\times \nonumber\\
 \int_0^{\infty}s s_{1z} r_p(s)\left(J_0(s \rho)-\dfrac{J_1(s \rho)}{s\rho}\right)\exp(is_{1z}z)ds.
%m'_{xx}=-a_2(s,\rho),\nonumber \\
%a_2=2\pi\left(j_0(s \rho)-\dfrac{j_1(s \rho)}{s\rho}\right).\nonumber 
\end{gather}

With this we need to compute $\partial_{x}G_{s, xx}$: 
$$
\dd_x J_0(s \rho)={k_0}\dd_{\rho}J_0(s \rho)=-{k_0}sJ_{1}(s\rho)
$$

$$
\dd_x \dfrac{J_1(s \rho)}{s\rho}=k_0s\dd_{s\rho}\dfrac{J_1(s \rho)}{s\rho}=-k_0s\dfrac{J_2(s \rho)}{(s\rho)}
$$
 
which gives us 
\begin{gather}
\dd_x G_{s, xx}(\rho, 0, z)=\frac{ik_1}{4\pi}{k_0}\int_0^{\infty}s^2 s_{1z} r_p(s)\times \\
\left(J_1(s \rho)-\dfrac{J_2(s \rho)}{(s\rho)^2}\right)\exp(is_{1z}z)ds.
%m'_{xx}=-a_2(s,\rho),\nonumber \\
%a_2=2\pi\left(j_0(s \rho)-\dfrac{j_1(s \rho)}{s\rho}\right).\nonumber 
\end{gather}

In order to compute the integral with help of  complex analysis, we first continue the integral bounds to $-\infty, +\infty$ using of the identity: 
$$
J_n(q)=\dfrac12(H_n^{(1)}(q)-(-1)^nH_n^{(1)}(-q)).%\Rightarrow j_1(z)=\dfrac12 (h_1^{(1)}(z)-h_1^{(1)}(-z))
$$

\begin{gather}
\dd_x G_{s, xx}(\rho,0 , z)=\frac{ik_1}{8\pi}{k_0}\times \nonumber \\ \int_{-\infty}^{\infty}\underbrace{s^2 s_{1z} r_p(s)\left(H_1^{(1)}(s \rho)-\dfrac{H_2^{(1)}(s \rho)}{(s\rho)}\right)\exp(is_{1z}z)}_{{I(s) }}ds. 
\end{gather}

Now, using Cauchy theorem we  finally evaluate this integral: 
\begin{gather}
\dd_xG_{s, xx}(\rho,0 , z)=\frac{ik_1}{8\pi}{k_0} {2 \pi i \mbox{Res}(I(s))|_{s=\tilde s}}=\nonumber \\-\frac{ k_1}{4}k_0 (\tilde s)^2 \tilde s_{1z} \left(H_1^{(1)}(\tilde s \rho)-\dfrac{H_2^{(1)}(\tilde s \rho)}{(\tilde s\rho)}\right)\times \\
\exp(i\tilde s_{1z}z) \mbox{Res}(r_p(s))|_{s=\tilde s},  \nonumber
\end{gather}
where $\tilde s=\sqrt{\veps_2\veps_1/(\veps_1+\veps_2)}$ is dimensionless wavevector of SPP mode.

Finally, computing the explicit expression for the residue and substituting the obtained results into Eq.(\ref{Force_X2}) one can get: 
\bea
F_x=\pi |p_x|^2 \Re\left[\dfrac{(\tilde k)^3 (\tilde k_{1z})^2\tilde k_{2z}}{k_0^2(\veps_1-\veps_2)}\left(H_1^{(1)}(\tilde k x)\right.\right. - \nonumber \\
\left.\left.\dfrac{H_2^{(1)}(\tilde k x)}{(\tilde k x)^2}\right)\exp(i\tilde k_{1z} z) \right].\nonumber
\ena

Here we use dimension variables denoting $\tilde k=\tilde s k_0$, and $\tilde k_{z1,z2}=\tilde s_{z1,z2} k_0$. We can go even further taking into account that $|H_1^{(1)}(\tilde k x)|\ll| {J_2(\tilde k x)}/(\tilde k x)^2|$: 
\begin{gather}
F_x\approx \pi |p_x|^2 \Re\left[\dfrac{(\tilde k)^3 (\tilde k_{1z})^2\tilde k_{2z}}{k_0^2(\veps_1-\veps_2)}H_1^{(1)}(\tilde k x)\exp(i\tilde k_{1z} z) \right].
\end{gather}

The case of low losses is of special interest. Then, the final expression for the force can be reduced to:  
   
\begin{gather}
F_x=\pi |p_x|^2 \dfrac{(k^*)^3 |k_{1z}^*|^2|k_{2z}^*|}{k_0^2(\veps_1-\veps_2)}Y_1(k^* r)\exp(-k_{1z}^*z),\nonumber
\end{gather}

where $Y_1(q)$ is the cylindrical Webber function. By expanding this  expression around the zeros $q_n$ of the Webber function $Y_1(q)\approx -Y_2(q_n)(q-q_n)$, one can find the expression for the stiffness at the $n$-th equilibrium position along $x$-axis  of the system (see Fig.~\ref{Fig3} a)): 

\begin{gather}
\label{AppStiff}
\kappa_n\approx \pi |p_x|^2 \dfrac{(\tilde k)^3 |\tilde k_{1z}|^2|\tilde k_{2z}|}{k_0^2(\veps_1-\veps_2)}Y_2(q_{2n-1})\exp(-|\tilde k_{1z}|z).
\end{gather}
Note that in the regime of SPP excitation without ohmic losses the SPP wavevector can be in the range from $k_0$ to $+\infty$ when $\veps_1+\veps_2\rightarrow 0$. With that the expression Eq.~(\ref{AppStiff}) goes to zero in both limiting cases:
  $$\kappa_n\xrightarrow[\tilde k\rightarrow 0, \infty]{}0,$$
  
  which implies that the stiffness reaches its maximum at some particular wavelength. This wavelength can be defined for each  given distance over substrate $z$. The maximal stiffness  can be achieved close to SPP resonance when SPP wavevector equals to $\tilde k\approx6/z$. Close to the frequency of SPP resonance when $\tilde{k}\rightarrow \infty$, the SPP becomes highly localized close to the interface $ |\tilde k_z|\gg k_0$.  Then,  the maximal stiffness can be expressed as: 
\bee
\label{AppMaxStiff}
\kappa_n\sim |p_x|^2 \left(\dfrac6z\right)^6\dfrac{1}{k_0^2\veps_1}Y_2(q_{2n+1})\exp(-6).
\ene

%=====================New section====================================================
\section{Dynamics simulation}
\label{App:dynamics}

We write down the equation of the Newton's equation  for the second particle
 $$m \frac{d^2}{dt^2}\ve{r}_2 = \ve{F}_2,$$
 where $\ve{F}_2$ is given by \eqref{eq:A_Force} and \eqref{eq:A_Field}. One can rewrite it as:
\begin{gather}
\frac{d^2}{dt^2} \ve{r}_2 = \frac{1}{2m} \sum_i\Re \bigg\{ p_{2i}^* \ve{\nabla} \bigg( E_{0i} + \frac{k^2}{\varepsilon_0} \sum_j G_{s,ij}(\ve{r}_2, \ve{r}_2) p_{2j} + 
%\right. \right.
\\
%\left. \left.
+ \frac{k^2}{\varepsilon_0} \sum_j G_{ij}(\ve{r}_2, \ve{r}_1) p_{1j}\bigg)\bigg\}, \ i,j=x,y,z \nonumber
\end{gather}
where $\widehat{G} = \widehat{G}_0 + \widehat{G}_s$. In order to decrease the numerical error during numerical simulations,  we apply the following natural scaling:
\begin{gather}
%	\begin{aligned}
	\bm{\xi} = \dfrac{\ve{r}}{a}, \quad \tau = \dfrac{t}{T}, \quad \widetilde{\ve{E}} = \dfrac{\ve{E}}{E_0}, \quad \widetilde{\widehat{G}} = a \widehat{G}, \\ 
\widetilde{\ve{k}} = a \ve{k}, \quad \widetilde{\ve{p}} = \dfrac{\ve{p}}{4\pi \varepsilon_0 a^3 E_0}, \quad \widetilde{\alpha} = \dfrac{\alpha}{4\pi \varepsilon_0 a^3},\nonumber \\
		 T=\sqrt{\dfrac{m}{2\pi\veps_0 a E_0^2}}.\nonumber
\end{gather}
After such substitutions we have
\begin{multline}
	\frac{d^2}{d\tau^2} \bm{\xi}_2 =  \sum_i \Re \bigg\{ \widetilde{p}^*_{2i} \frac{\dd}{\dd \bm{\xi}} \bigg( \widetilde{E}_{0i} + \widetilde{k}^2 \sum_j \widetilde{G}^s_{ij}(\bm{\xi}_2,\bm{\xi}_2) \widetilde{p}_{2j} + 
	%\right. \right.
	\\
	%\left. \left. 
	+ \widetilde{k}^2 \sum_j  	\widetilde{G}_{ij}(\bm{\xi}_2,\bm{\xi}_1) \widetilde{p}_{1j}\bigg) \bigg\}.
	\label{eq:dimless}
\end{multline}
We also include the viscosity of the environment by adding the damping factor $\gamma$: 

\begin{equation}
		\frac{d^2}{d\tau^2} \bm{\xi}_2 = \widetilde{\ve{F}}_2(\bm{\xi}_1,\bm{\xi}_2) - \gamma \frac{d}{d\tau}  \bm{\xi}_2,
		\label{eq:to_solve}
\end{equation}
where $\widetilde{\ve{F}}_2$ is given by the r.h.s. of Eq.~\eqref{eq:dimless}.
Expression  \eqref{eq:to_solve} was a target for the numerical simulation. As a good compromise between stability and computational complexity the Rungeâ-Kutta of fourth-order method was applied. Due to the fact that the motion along $z$ axis is fixed, we have plane symmetry, which simplifies the force function to $\widetilde{\ve{F}}_2(\bm{\xi}_1,\bm{\xi}_2) = \widetilde{\ve{F}}_2(\bm{\xi}_1-\bm{\xi}_2)$. % so it is sufficient to tabulate the force field on a grid at $(x,y)$ plane at the essential hight $z_0$ with sufficient accuracy for afterwards interpolation. Such move significantly decrease the computational time.  
 
%

%\section{Conclusion}
%=====================New section====================================================
\section{Conservative vs non-conservative force components}
\label{App:conserve}
\begin{figure}[h!]
\includegraphics[width=1\columnwidth]{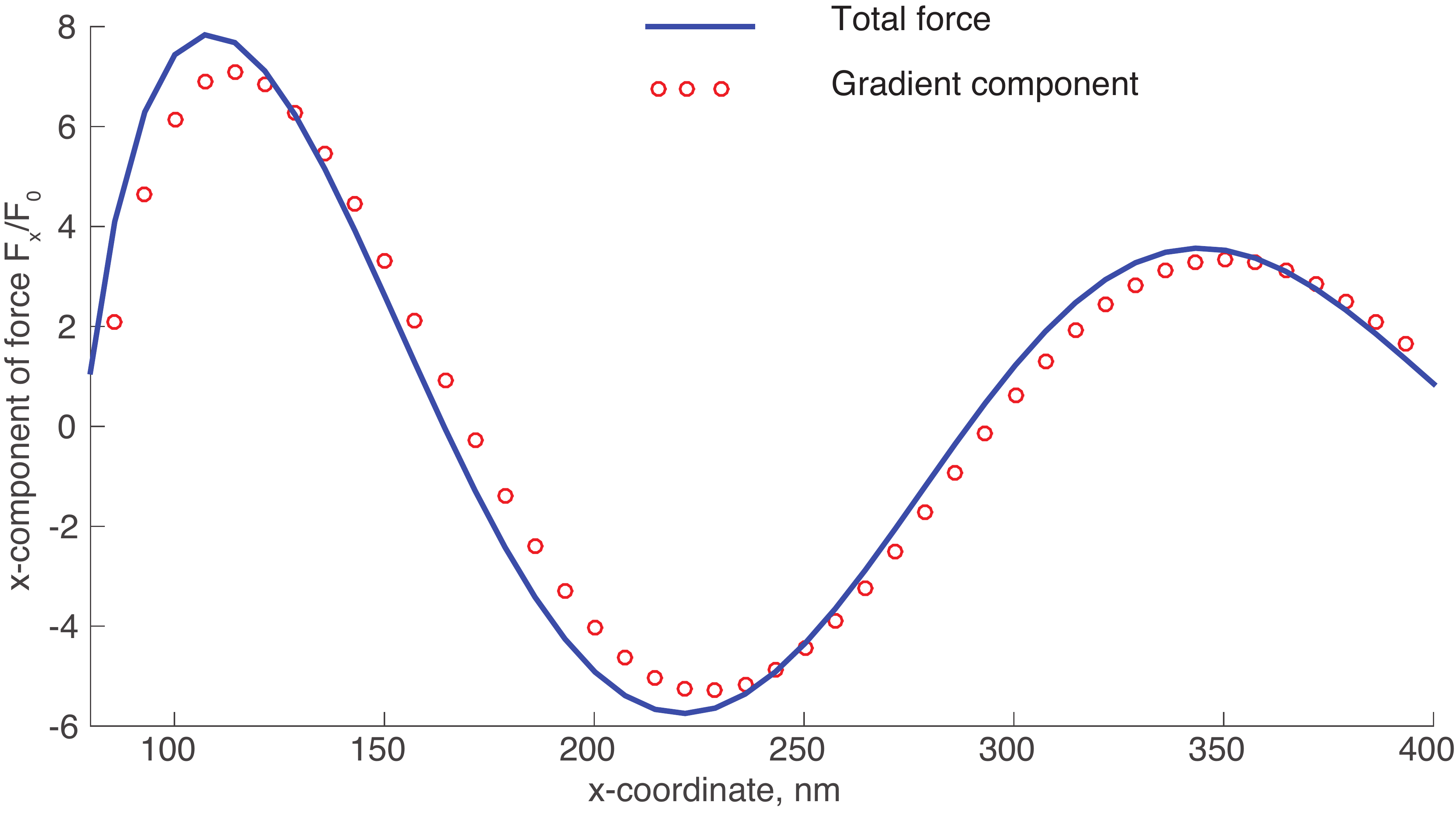}
\caption{ The total force (blue solid line) and the conservative (red circles) components of the optical force are shown for different wavelengths. The parameters of the calculation are the same as in Fig.~\ref{Fig2}.}%Inset: the binding length versus the  vacuum wavelength, and SPP period.  }
\label{FigApp}
\end{figure}

Here we present the results of calculation of total optical force and the conservative component only. By excluding imaginary part of polarizability one can isolate the conservative force only \cite{Novotny2012}. The result are shown in Fig.~\ref{FigApp}. One can see that for the considered set of parameters the conservative force strongly dominates over non-conservative, which is the the difference between the total and conservative forces.  
%===============================FIGURE====================================================================
%===============================FIGURE====================================================================

\bibliography{Rfrncs_spp_bind1}

\end{document}